\documentstyle[12pt,epsfig]{article}
\begin{document}
\begin{flushright}
DTP/00/02  \\
January 2000\\
\end{flushright}
\begin{center}
{\Large \bf The interplay between perturbative QCD and power corrections:
the description  of   scaling  or automodelling limit violation
in deep-inelastic scattering
}
\bigskip

{\large A.L.~Kataev$^{a,}$\footnote{Supported by UK Royal Society 
and in part by the Russian 
Foundation of Basic Research, Grant N 99-01-00091}, 
G.~Parente$^{b,}$\footnote{Supported by CICYT (AEN96-1773) and Xunta 
de Galicia (Xuga-20602B98)} A.V.~Sidorov$^{c,}$\footnote{Supported in part 
by the Russian Foundation of Basic Research, Grant N 99-01-00091}}
\date{}
\smallskip

{\it (a) Centre for Particle Theory of the University of Durham,\\ 
DH1 3LE, Durham, United Kingdom and \\
Institute for Nuclear Research of the Academy of Sciences of Russia,\\
117312 Moscow, Russian Federation\footnote{Permanent address}\\
(b) Department of Particle Physics, University of Santiago de Compostela,\\ 
15706 Santiago de Compostela, Spain \\
(c) Bogolyubov Laboratory of Theoretical Physics,
Joint Institute for Nuclear Research, 141980 Dubna, Russian Federation
}

\hspace{3cm}

\end{center}

\begin{abstract}
The summary of  the results of our next-to-next-to-leading  fits of the 
Tevatron experimental data for $xF_3$ structure function of the $\nu N$
deep-inelastic scattering is given. The special attention is paid to 
the extraction of twist-4  contributions and demonstration 
of the  interplay between these effects and higher 
order perturbative QCD corrections. The factorization and renormalization 
scale uncertainties of the results obtained are analysed. 
\end{abstract}

\hspace{3cm}

\begin{center}

{\it Contributed to Bogolyubov Conference \\``Problems of Theoretical 
and Mathematical Physics"\\ Moscow-Dubna-Kyiv, 27 September - 6 October 1999} 

\end{center}
\newpage
{\bf 1.}~The study of deep-inelastic scattering (DIS) 
processes has a rather long and inspiring history. One of the first 
realizations that the analysis of $\nu N$ DIS could play an important role 
in investigations of the properties of the nucleon came 
 in Ref.\cite{Markov}. The fundamental concept of scaling of 
 DIS structure functions (SFs) \cite{Bj2} has lead to many subsequent 
 investigations. Other important stages in the 
development of both theoretical and experimental studies 
of various characteristics of DIS processes in this productive 
period were reviewed in detail  recently \cite{Drell}. 
In particular, it was stressed that after the experimental 
confirmation of scaling  and indications of the existence 
of point-like constituents of the nucleon, the more rigorous theoretical 
explanation of the behaviour of DIS form factors 
came onto the agenda. A series of 
works by N. N. Bogolyubov and coauthors \cite{BVT}, were devoted to 
the development of the new method, which made it possible to analyse the 
asymptotics of the form factors 
of  $eN$ DIS  using the Jost-Lehmann-Dyson 
integral representation, 
and explain the property 
of scaling (or as called it by the authors of Ref.\cite{BVT} ``automodelling") 
behaviour of the corresponding SFs
in the framework of general principles of local quantum field 
theory \cite{BSh}.

We now know that this property is true only in the asymptotic 
regime and that  it is violated within the framework of QCD 
(see e.g. the extensive 
discussions in a number of books on the subject \cite{Books}).
Indeed, the theory of QCD predicts that scaling or automodelling 
behaviour of SFs is violated by the logarithmically 
decreasing perturbative QCD contributions 
to the leading twist operators. However, in the intermediate 
and low $Q^2$ regime the higher twist operators, which 
give rise to  scaling violations of the form $1/Q^2$, 
$1/Q^4$, etc., might also be  important \cite{HT1,HT2}. 
Indeed, the  NLO DGLAP fits \cite{VM} 
of the BCDMS data of DIS of charged leptons on nucleons \cite{BCDMS} 
and reanalysed SLAC $eN$ data \cite{SLAC} resulted in the detection of 
the signals from the twist-4 contributions.

During the last few years there has been
considerable progress in modeling these effects
with the help of the infrared renormalon (IRR) approach (for the review 
see Ref.\cite{IRR})
and  the dispersive method \cite{disp}  (see also Ref.\cite{disp1}).
Using these methods the authors of Ref.\cite{DW} 
explained the  behaviour 
of the twist-4 contributions to the $F_2$ SF observed in Ref.\cite{VM} 
and constructed a model 
for the similar power-suppressed corrections to $xF_3$ SF.  
In view of this  it became important to check the predictions of Ref.\cite{DW}
and to study the possibility of extracting higher-twist contributions 
from the new more precise 
experimental data for $\nu N$ DIS, obtained by the CCFR 
collaboration at Fermilab Tevatron \cite{CCFR}, and 
also to exploit the  
considerable progress in calculations of the 
perturbative QCD corrections to characteristics of DIS, achieved in the 
last decade.

Indeed, the analytic expressions for the  
next-to-next-to-leading order (NNLO) perturbative QCD corrections 
to the coefficient functions of
SFs $F_2$ \cite{VZ} and $xF_3$ \cite{ZV} are now known. 
Moreover, the
expressions for the NNLO corrections to the anomalous dimensions of non-singlet (NS)
even Mellin moments of $F_2$ SF  with $n=2,4,6,8,10$   and for the 
N$^3$LO corrections 
to the coefficient functions of  these moments  
are also available \cite{LRV}. 
In this report we will summarize the results of the series of the 
works of Refs.\cite{KKPS1}-\cite{KPS1}, devoted to 
the analysis of the CCFR data at  NNLO, which has the aim 
to  determine  
the NNLO value of the QCD coupling constant $\alpha_s(M_Z)$ and 
to extract the effects 
of the twist-4 contributions to SF $xF_3$ \cite{KKPS2,KPS1}.
In particular, we will concentrate  on the discussion 
of the factorization and renormalization
scale uncertainties of the results obtained.

{\bf 2.}~ Our analysis of Refs.\cite{KKPS1}-
\cite{KPS1} is based on reconstruction of the SF $xF_3$ from 
its Mellin moments $M_n(Q^2)=\int_0^1x^{n-1}F_3(x,Q^2)dx$ using the Jacobi polynomials method, proposed 
in Ref.\cite{PS} and further developed in the works of Ref.\cite{Jacobi}.
Within this framework one has:
\begin{equation}
xF_3(x,Q^2)=x^{\alpha}(1-x)^{\beta} \sum_{n=0}^{N_{max}}\Theta_{n}^{\alpha,\beta}(x)
\sum_{j=0}^{n}c_j^{(n)}(\alpha,\beta)M_{j+2}(Q^2)
\end{equation}
where  
$\Theta_{n}^{\alpha,\beta}$ are the Jacobi polynomials,  
$c_j^{(n)}(\alpha,\beta)$ are combinatorial coefficients given in 
terms of Euler $\Gamma$-functions of the $\alpha$ and $\beta$ weight parameters.
In view of the reasons discussed in Ref.\cite{KPS1}, they were fixed 
to 0.7 and 3 respectively. The QCD evolution of the moments 
is defined by the solution of the corresponding renormalization group equation:
\begin{equation}
\frac{M_n(Q^2)}{M_n(Q_0^2)}=exp \bigg[-\int_{A_s(Q_0^2)}^{A_s(Q^2)}
\frac{\gamma_{NS}^{(n)}(x)}{\beta(x)}dx\bigg]
\frac{C_{NS}^{(n)}(A_s(Q^2))}{C_{NS}^{(n)}(A_s(Q_0^2)}
\end{equation}
The QCD running coupling constant  enters this equation 
through  $A_s(Q^2)=\alpha_s(Q^2)/(4\pi)$ and is defined as the expansion 
in terms of inverse powers of $ln(Q^2/\Lambda_{\overline{MS}}^{(4)~2})$. 
For the initial scale $Q_0^2$, from which the evolution is started, the 
moments in Eq.(2) were parametrized as 
$M_n(Q_0^2)=\int_0^{1}x^{n-2}A(Q_0^2)x^{b(Q_0^2)}(1-x)^{c(Q_0^2)}
(1+\gamma(Q_0^2) x)dx$. 
In the process of our analysis we took into account both target 
mass corrections and twist-4 contributions.
The latter  were modeled using the  IRR approach   
as $M_n^{IRR}=C(n)M_n(Q^2)A_2^{'}/Q^2$ \cite{DW} and by adding into the r.h.s. 
of Eq.(1) the term $h(x)/Q^2$ with $h(x)$ considered as a free parameter 
for each $x$-bin of the  experimental data. 

For  arbitrary factorization and renormalization scales
the NNLO expression 
for the NS Mellin moments reads:
\begin{equation}
M_n(Q^2) \sim (A_s(Q^2 k_F))^{a}\times \overline{AD}(n,A_s(Q^2k_F))
\times C_{NS}^{(n)}(A_s(Q^2 k_R))
\end{equation}
where $a=\gamma_{NS}^{(0)}/(2\beta_0)$, 
$\overline{AD}=1+\bigg[p(n)+ak^F_1\bigg]A_s(Q^2k_F)+\bigg[q(n)+p(n)(a+1)k_1^F
+ (\beta_1/\beta_0)ak^F_1+a(a+1)(k^F_1)^2/2\bigg]
A_s^2(Q^2k_F)$ and $C_{NS}^{(n)}=1+C^{(1)}(n)A_s(Q^2k_R)+\bigg[C^{(2)}(n)+
C^{(1)}(n)k_1^R\bigg]A_s^2(Q^2k_R)$. Here $\gamma_{NS}^{(0)}$, $\beta_0$ and 
$\beta_1$ are the scheme-independent coefficients of the 
anomalous dimension function $\gamma_{NS}(x)$ and QCD $\beta$-function
$\beta(x)$, $p(n)$ and $q(n)$-terms are expressed through 
the  NLO and NNLO  coefficients of $\gamma_{NS}(x)$ and $\beta(x)$ 
via equations, given in 
Refs.\cite{KKPS1,KPS1}. Within the $\overline{MS}$-like schemes 
the factorization and renormalization scale ambiguities are parameterized 
by the terms $k_1^F=\beta_0ln(k_F)$ and $k_1^R=\beta_0 ln(k_R)$, where 
$k_F$ ($k_R$) is the ratio of the factorization (renormalization) scale and 
the scale of the $\overline{MS}$-scheme. Following the analysis of Ref.\cite{NV} 
we  
take $k_R=k_F=k$,  fixing identically the factorization scale and the 
renormalization scale. We performed our fits for the case of $k=1$ 
(namely, in the pure $\overline{MS}$-scheme) and then determine 
the scale uncertainties of  $\Lambda_{\overline{MS}}^{(4)}$, 
the twist-4 parameter $A_2^{'}$ and the  $x$-shape 
of $h(x)$ by 
choosing 
$k=1/4$ and $k=4$ and repeating the fits for these two cases. 

{\bf 3.}~ In the process of our analysis of CCFR'97 
data we applied the same kinematic cuts as in Ref.\cite{CCFR}, 
namely $Q^2>5~GeV^2$, $x<0.7$ and $W^2>10~GeV^2$. We started the QCD 
evolution from the 
initial scale $Q_0^2=20~GeV^2$, which we consider as more appropriate 
from the point of view of stability  of the NLO and  NNLO results for 
$\Lambda_{\overline{MS}}^{(4)}$ due to variation of the initial scale 
\cite{KPS1}. 
In order to estimate the uncertainties of the NNLO results, we 
also performed the N$^3$LO fits with the help of the expanded Pad\'e 
approximations technique (for the detailed discussions see Ref.\cite{KPS1}).
The results  are presented in Table 1. 
\begin{center}
\begin{tabular}{||r|c|c|c|}
\hline
                                   &
$\Lambda_{\overline{MS}}^{(4)}$ (MeV)  &                
$ A_2^{'}$ (GeV$^2$)                   &
$\chi^2$/points                        \\
\hline
LO & 264$\pm$37   & --             & 113.1/86  \\
   & 433$\pm$53   & -0.33$\pm$0.06 &  83.1/86  \\
   & 331$\pm$162  & h(x) in Fig.1  &  66.3/86   \\
\hline
NLO & 339$\pm$42  & --              &  87.6/86   \\
    &  369$\pm$39 &  -0.12$\pm$0.06 &  82.3/86   \\
    & 440$\pm$183 &  h(x) in Fig.1  &  65.8/86   \\
\hline
NNLO & 326$\pm$35  & --              &  77.0/86   \\
     & 327$\pm$35  & -0.01$\pm$0.05  &  76.9/86   \\
     & 372$\pm$133 & h(x) in Fig.1   &   65.0/86 \\
\hline
N$^3$LO & 332$\pm$28 & --            &  76.9/86 \\
Pade    & 333$\pm$27 & -0.04$\pm$0.05 & 76.3/86  \\
        & 371$\pm$127&  h(x) in Fig.1 & 64.8/86  \\
\hline
\end{tabular}
\end{center}
{{\bf Table 1.} The results of the fits of CCFR'97 data with the 
cut $Q^2>5~GeV^2$.} 
\vspace{0.5cm}

At  NLO the value for $\Lambda_{\overline{MS}}^{(4)}$ is in good agreement 
with the NLO result $\Lambda_{\overline{MS}}^{(4)}=337\pm28~MeV$, obtained 
by the CCFR collaboration with the help of DGLAP NLO analysis of both $F_2$ 
and $xF_3$ SFs data in the case when HT-corrections were neglected \cite{CCFR}.
The obtained NLO value of the IRR-model parameter $A_2^{'}$ is in agreement 
with the estimates of Ref.\cite{DW} and of Ref.\cite{Maul} especially.
However, at  NNLO a  significant decrease of the 
magnitude of the parameter $A_2^{'}$ is observed. In view of this 
the results for $\Lambda_{\overline{MS}}^{(4)}$ obtained at the NNLO 
without HT corrections and with IRR-model of twist-4 term almost coincide. 
A similar tendency was observed in the process of the N$^3$LO Pad\'e fits.
To study this feature in more detail we extracted the $x$-shape 
of the model-independent function $h(x)$  (see Fig.1) and analysed the 
factorization/renormalization scale uncertainties of the outcomes of our fits 
\cite{KPS1}. The corresponding results are presented in Table 2 
where $\Delta_k$ is defined as $\Delta_k=\Lambda_{\overline{MS}}^{(4)}(k)-
\Lambda_{\overline{MS}}^{(4)}(k=1)$. The related $x$-shapes of $h(x)$ 
are presented in Fig.2.
\begin{center}
\begin{tabular}{||r|c|c|c|c|}
\hline
Order & $k$    & $\Delta_k$ (MeV)  & $A_2^{'}$ (GeV$^2$) & $\chi^2$/points \\  
\hline
NLO   & 4      & 116               & --                  & 99.1/86        \\
      & 4      & 213               & -0.22$\pm$0.006     & 84.2/86         \\
      & 1/4    & -61               & --                  & 80.4/86         \\
      & 1/4    & -99               & +0.02$\pm$0.005     & 80.2/86         \\
\hline
NNLO  & 4      & 35                & --                  & 83.5/86         \\
      & 4      & 66                & -0.11$\pm$0.06      & 83.5/86         \\
      & 1/4    & -51               & --                  & 87.3/86         \\
      & 1/4    & -45               & +0.09$\pm$0.05      & 84.5/86          \\
\hline
\end{tabular}
\end{center}
{{\bf Table 2.} The results of NLO and NNLO fits of CCFR'97 data for 
different values of factorization/renormalization scales.}

\begin{center}
\begin{figure}
\psfig{figure=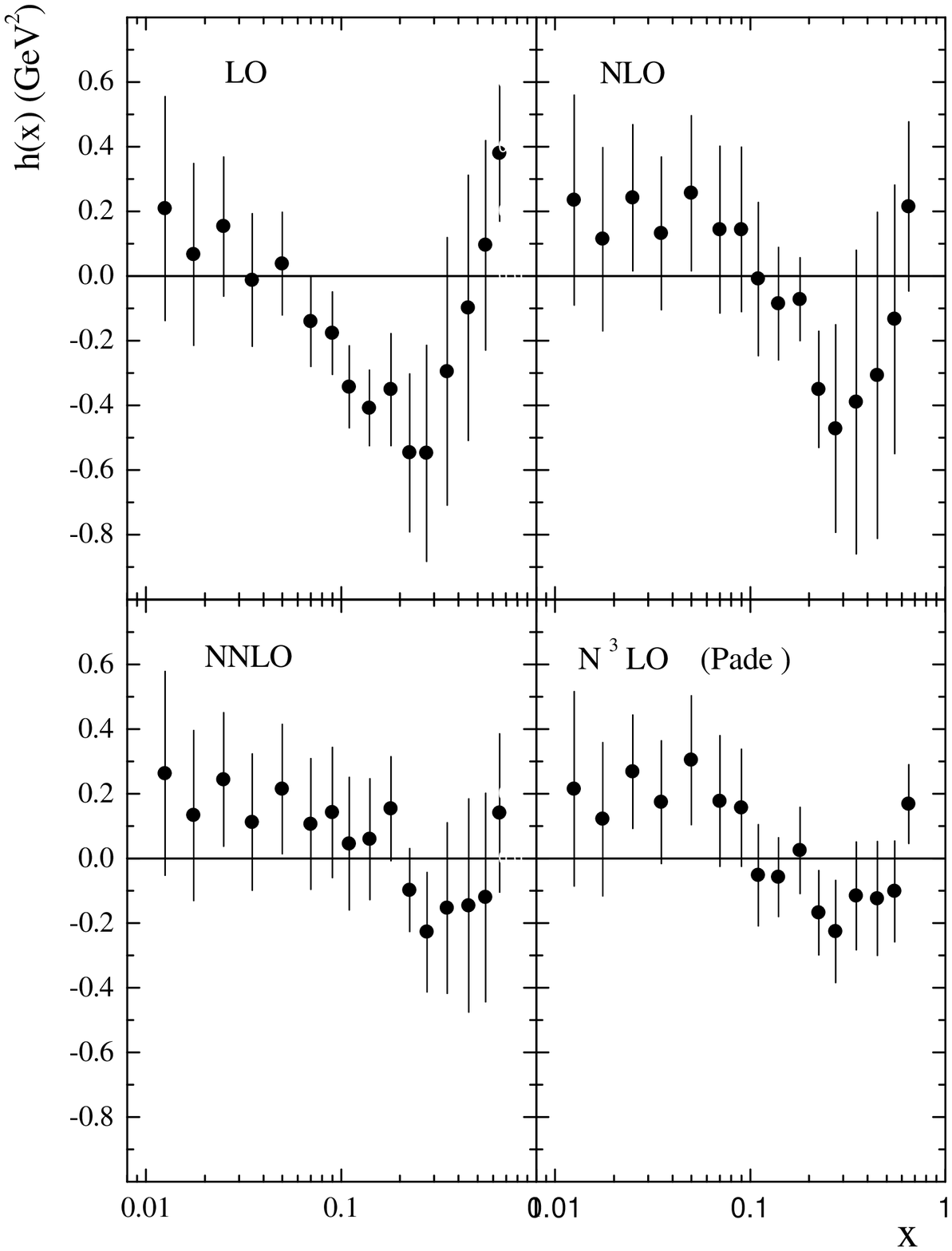, height=6.5cm,width=10cm}
\caption{$h(x)$ extracted from CCFR'97 data}
\psfig{figure=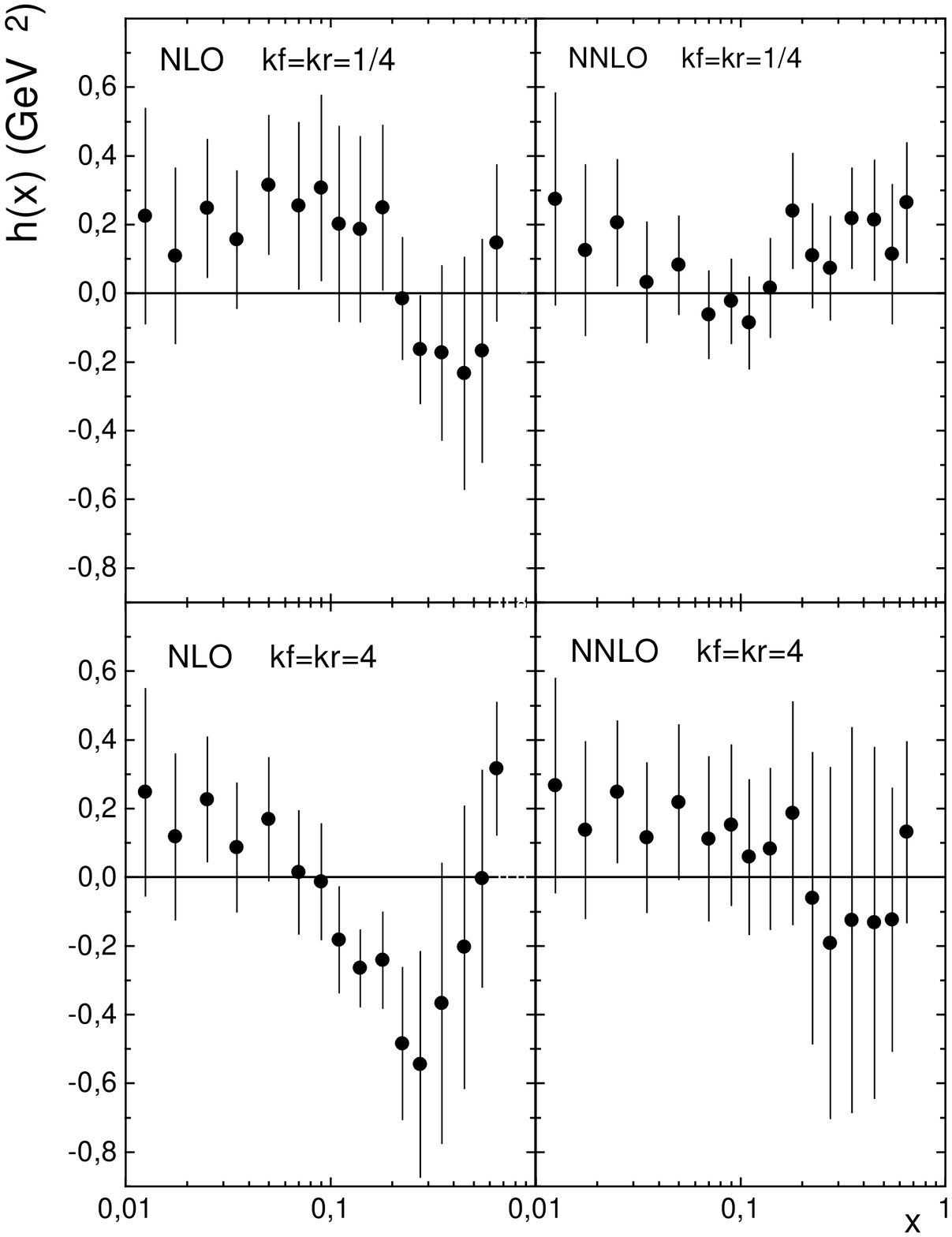, height=6.5cm, width=10cm} 
\caption{ Scale dependence 
of $h(x)$}
\end{figure}
\end{center}

{\bf 4.} We will concentrate first on  
discussing  the presented behaviour of the twist-4 
parameter $h(x)$ of $xF_3$ SF, presented in Figs.1,2. 
In the case of $k=1$, namely in the pure 
$\overline{MS}$-scheme, 
$x$-shape of $h(x)$
obtained from the LO and NLO analysis of Refs.\cite{KKPS2,KPS1} 
is in 
agreement 
with the IRR-model predictions of Ref.\cite{DW}. Note also that the 
combination of the quark counting rules \cite{qcr} with the 
results  of Ref.\cite{HT1} 
predict the following $x$-form of $h(x)$: $h(x)\sim A_2^{'}
(1-x)^2$. Taking into account the  negative  
values of $A_2^{'}$, obtained in the process of 
our LO and NLO fits (see  Table 1), we  conclude that the related  
behaviour of $h(x)$ is in qualitative agreement with these
predictions. 

At the NNLO  the situation is more intriguing. Indeed, though a certain 
indication    
of the twist-4 term survives even 
at this level, the NNLO part of Fig.1 demonstrates that its 
extracted $x$-shape  starts 
to deviate 
both from the IRR prediction  of Ref.\cite{DW} and from 
the  quark-parton model picture, mentioned above. Notice also that 
within the statistical error bars the NNLO value of $A_2^{'}$  
is indistinguishable from zero. 
These conclusions are confirmed by the studies of the 
factorization/renormalization scale dependence of the 
NLO and NNLO outcomes of the fits \cite{KPS1}.

Indeed, it is known that the variation of the related scales 
is simulating in part the effects of the higher-order 
perturbative QCD corrections. In view of this the NLO (NNLO) 
results, obtained in the case of $k=1/4$ (see Table 2 and 
Fig.2 in particular), are almost identical to the NNLO (Pad\'e motivated 
N$^3$LO) extractions of $h(x)$ and of the IRR model parameter $A_2^{'}$ 
from the fits with $k=1$ (see Fig.1 and Table 1). Thus, we conclude, 
that as the result of analysis of the CCFR'97 data 
the NNLO and beyond we observe the minimization  
of the 
twist-4 contributions to $xF_3$ SF. This feature 
is related to the interplay between NNLO perturbative 
QCD  and twist-4 $1/Q^2$ corrections. 
The recent studies of the scale-dependence 
of the  
NLO DGLAP extraction of the twist-4 terms from different  recent  
DIS experimental data  \cite{AK} are supporting the foundations 
of Refs.\cite{KKPS2,KPS1}. This means that the higher-twist 
parameters cannot be defined independently of the effects of perturbation 
theory and that the NNLO corrections  can mimick the contributions 
of higher twists  \cite{St} provided the experimental data is not precise 
enough for the clear separation of the nonperturbative 
from perturbative effects. Thus, it is highly desirable to have 
new experimental data for $xF_3$ SF, which are more precise than the 
ones given by the CCFR collaboration.   
    
In conclusion we present also the NLO and NNLO values of $\alpha_s(M_Z)$, 
obtained by us in Ref.\cite{KPS1} from the fits of CCFR'97 data for $xF_3$ 
SF with twist-4 terms modeled through the IRR approach :
\begin{eqnarray}
NLO~~~\alpha_s(M_Z) &=& 0.120 \pm 0.003 (stat) \pm 0.005 (syst)^{+ 0.009}
_{-0.007} \\
\nonumber 
NNLO ~~\alpha_s(M_Z) &=& 0.118 \pm 0.003 (stat) \pm 0.005 (syst) \pm 0.003
\end{eqnarray}
The systematical uncertainties in these results are determined by 
the systematical  uncertainties of the CCFR'97 data and the 
theoretical errors are fixed 
from the numbers for $\Delta_k$ (see Table 2), which reflect 
the factorization/renormalization scale uncertainties of the values 
of $\Lambda_{\overline{MS}}^{(4)}$. The 
incorporation into the $\overline{MS}$-matching formula 
\cite{match} of the proposals 
of Ref.\cite{MOM} 
for estimates of the ambiguities due to smooth transition to the world with 
$f=5$ numbers of active flavours was also taken into account. 
The  theoretical 
uncertainties presented  are in agreement with the ones, obtained in Ref.\cite{NV}, while 
the NNLO value of $\alpha_s(M_Z)$ is in agreement with another  
NNLO result  $\alpha_s(M_Z)=0.1172 \pm 0.0024$, which was obtained 
from the analysis of SLAC, BCDMS, E665 and HERA data for $F_2$ SF with 
the help of the Bernstein polynomial technique \cite{SYnd}.
It might be of interest to verify the theoretical errors
of these two available phenomenological NNLO analysis using 
different variants of fixing scheme-dependence ambiguities. 
The first steps towards the  analysis of this problem 
are already made \cite{Ch}.

{\bf Acknowledgements.} We are grateful to C.J. Maxwell for careful 
reading of the manuscript.


\newpage
\begin{thebibliography}{99}
\bibitem{Markov} M. A. Markov, {\it The Neutrino [in English]}, 
Preprint JINR, D-1269 (1963), published as the book 
by Fizmatgiz (1964), translated to Japanese.
\bibitem{Bj2} J. D. Bjorken, {\it Phys. Rev.} {\bf 179} (1969) 1547. 
\bibitem{Drell} R. J. Jaffe, Preprint MIT-CTP-2791; hep-ph/9811327;\\
C. H. Llewellyn Smith, Preprint CERN-DG/98-3534; hep-ph/9812301.
\bibitem{BVT}
N. N. Bogolyubov, V. S. Vladimirov, A. N. Tavkhelidze, {\it Theor. Math. 
Phys.} {\bf 12} (1972) 3, 305.
\bibitem{BSh} N. N. Bogolyubov, D.V. Shirkov, {\it Introduction to the 
Theory of Quantum Fields}, Nauka, Moskow (1973, 1976, 1986);
{\it English transl.} Wiley, NY (1959,1980).
\bibitem{Books}
F. J. Yndur\'ain, {\it Quantum chromodynamics: an introduction to the 
theory of quarks and gluons}, Springer  Verlag (1983);\\
B.L. Ioffe, V.A. Khoze, L.N. Lipatov, {\it Hard processes. v1: Phenomenology, 
quark parton model}, {\it English transl.} North Holland (1984)\\
G. Altarelli, {\it The development of perturbative QCD}, World Scient. (1994);\\
R. K. Ellis, W.J. Stirling, B.R. Webber, {\it QCD and colliders}, Cambridge 
Univ. Press (1996).
\bibitem{HT1}
E. Berger, S.J. Brodsky, {\it Phys. Rev. Lett.} {\bf 42} (1979) 940;\\
J. Gunion, P. Nason, R. Blankenbecler, 
{\it Phys. Rev.} {\bf D29} (1984) 2491.
\bibitem{HT2}
R.L. Jaffe, M. Soldate, {\it Phys. Rev.} {\bf D26} (1982) 49;\\
S.P. Luttrell, S. Wada, B.R. Webber, {\it Nucl. Phys.} {\bf B188} 
(1981) 219;\\
E.V. Shuryak, A.I. Vainshtein, {\it Nucl. Phys.} {\bf B199} (1982) 451.
\bibitem{VM}
M. Virchaux, A.Milsztain, {\it Phys. Lett.} {\bf B274} (1992) 221.
\bibitem{BCDMS}
BCDMS Collab., C. Benvenuti et al., {\it Phys. Lett.} {\bf B195} (1987) 97;
{\it Phys. Lett.} {\bf B232} (1989) 490. 
\bibitem{SLAC}
L.W. Whitlow, Ph. D. Thesis; SLAC-0357 (1990).
\bibitem{IRR}
M. Beneke, {\it Phys. Rept.} {\bf 317} (1999) 1. 
\bibitem{disp}
Yu.L.Dokshitzer, G.Marchesini, B.R.Webber, {\it Nucl. Phys.} 
{\bf B469} (1996) 93.
\bibitem{disp1}
D.V. Shirkov, I.L. Solovtsov, {\it Phys. Rev. Lett.} {\bf 79} (1997) 
1209;
I.L. Solovtsov, D.V. Shirkov, {\it Theor. Math. Phys.} 
{\bf 120} (1999) 482 (hep-ph/9909305).
\bibitem{DW}
M. Dasgupta, B.R. Webber, {\it Phys. Lett.} {\bf B382} (1996) 273.
\bibitem{CCFR}
CCFR-NuTeV Collab., W.G. Seligman et al., {\it Phys. Rev. Lett.} 
{\bf 79} (1997) 213.
\bibitem{VZ}
W.L. van Neerven, E.B. Zijlstra, {\it Phys. Lett.} {\bf B272} (1991)
127; ibid. {\bf B273} (1991) 476; {\it Nucl. Phys.} {\bf B383} (1992) 525.
\bibitem{ZV} 
E.B. Zijlstra, W.L. van Neerven, {\it Phys. Lett.} {\bf B297} (1992) 377;
{\it Nucl. Phys.} {\bf B417} (1994) 61.
\bibitem{LRV}
S.A. Larin, T. van Ritbergen, J.A.M. Vermaseren, {\it Nucl. Phys.} 
{\bf B427} (1994) 41;
S.A. Larin, P. Nogueira, T. van Ritbergen, J.A.M. Vermaseren, 
{\it Nucl. Phys.} {\bf B492} (1997) 338.
\bibitem{KKPS1} 
A.L.Kataev, A.V.Kotikov, G.Parente, A.V.Sidorov, 
{\it Phys. Lett.} {\bf B388} (1996) 179.
\bibitem{KKPS2}
A.L.Kataev, A.V.Kotikov, G.Parente, A.V.Sidorov, 
{\it Phys. Lett.} {\bf B417} (1998) 374.
\bibitem{KPS1}
A.L.Kataev, G. Parente, A.V. Sidorov, Preprint IC/99/51 (1999) 
and hep-ph/9905310; to be published in {\it Nucl. Phys.} {\bf B}.
\bibitem{PS}
G. Parisi, N. Sourlas, {\it Nucl. Phys.} {\bf B151} (1979) 421;\\
I.S. Barker, C.B. Langensiepen, G. Shaw, {\it Nucl. Phys.} {\bf B186} 
(1981) 61.
\bibitem{Jacobi}
J. Ch\'yla, J. Ramez, {\it Z. Phys.} {\bf C31} (1986) 151;\\
V.G. Krivokhizhin et al., {\it Z. Phys.} {\bf C36} (1987) 51; 
ibid. {\bf C48} (1990) 347.
\bibitem{NV} 
W.L. van Neerven, A. Vogt, preprint INLO-PUB 14/99 (1999) 
(hep-ph/9907472), to be published in {\it Nucl. Phys.} {\bf B}.
\bibitem{Maul} M. Maul, E. Stein, A. Sch\"afer, L. Mankiewicz, 
{\it Phys. Lett.} {\bf B401} (1997) 100.
\bibitem{qcr}
V.A. Matveev, R.M. Muradyan, A.N. Tavkhelidze, {\it Lett. Nuov. Cim.} 
{\bf 7} (1973) 719;
S.J. Brodsky, G.R. Farrar, {\it Phys. Rev. Lett.} {\bf 31} (1973) 1153.
\bibitem{AK}
S.I. Alekhin, hep-ph/9907350;
S.I. Alekhin, A.L. Kataev, hep-ph/9908349, 
{\it Nucl. Phys.} {\bf A} (in press).
\bibitem{St}
G. Sterman, hep-ph/9905548 and talk at the QCD-Euronet Workshop, Florence, 
Italy, October 1999, quoted  by Yu. L. Dokshitzer, hep-ph/9911299.
\bibitem{match}
W. Bernreuther, W.Wetzel, {\it Nucl. Phys.} {\bf B197} (1982) 288; 
Err. ibid. {\bf B513} (1998) 758; S.A. Larin, T. van Ritbergen, J.A.M. 
Vermaseren, {\it Nucl. Phys.} {\bf B438} (1995) 278; K.G. Chetyrkin, 
B.A. Kniehl, M.Steinhauser, {\it Phys.Rev.Lett.} {\bf 79} (1997) 2184.
\bibitem{MOM} 
J. Bl\"umlein, W.L. van Neerven, {\it Phys. Lett.} {\bf B450} 
(1999) 417.
\bibitem{SYnd} J. Santiago, F.J. Yndur\'ain, preprint FTUAM 99-8 
(hep-ph/9904344).
\bibitem{Ch} 
C.J. Maxwell, A. Mirjalili, work in progress (private communication 
to A.L.K.).
\end{thebibliography}
\end{document}